\documentclass[12pt]{article}
\usepackage{amsmath}
\usepackage{cases}
\usepackage{fullpage}
\usepackage{amsfonts}
\usepackage{graphicx}
\usepackage{mathrsfs}
\usepackage{amssymb}
\usepackage{float}
\newtheorem{pro}{Proposition}
\newtheorem{thm}{Theorem}

\def\remark#1{\noindent{\bf Remark #1\ }}

\def\var{\mathrm {var}}

\def\cor{\mathrm {cor}}
\newcommand{\bm}{\boldsymbol}

\def\tr{\mathrm {tr}}
\def\U{{\bf U}}
\def\A{{\bf A}}

\def\D{{\bf D}}
\def\R{{\bf R}}
\def\z{{\bf z}}
\def\Z{{\bf Z}}
\def\G{{\bf \Gamma}}

\def\I{{\bf I}}

\def\X{{\bf X}}

\def\Y{{\bf Y}}

\def\tr{\mathrm {tr}}

\def\bb{{\bm\beta}}
\def\bms{{\bm\Sigma}}
\def\diag{\mathrm {diag}}

\def\cp{\mathop{\rightarrow}\limits^{p}}
\def\cd{\mathop{\rightarrow}\limits^{d}}

\title{\bf Scalar-Invariant Test for High-Dimensional Regression Coefficients}
\author{Long Feng\footnote{{Corresponding author. Email: flnankai@126.com}}\\
{\em\small Northeast Normal University}}
\date{}
\begin{document}
\maketitle
\begin{abstract} \baselineskip 18pt
This article is concerned with simultaneous tests on linear regression coefficients in high-dimensional settings. When the dimensionality is larger than the sample size, the classic $F$-test is not applicable since the sample covariance matrix is not invertible. In order to overcome this issue, both Goeman, Finos and van Houwelingen (2011) and Zhong and Chen (2011) proposed their test procedures after excluding the $(\X^{'}\X)^{-1}$ term in $F$-statistics. However, both these two test are not invariant under the group of scalar transformations. In order to treat those variables in a `fair' way, we proposed a new test statistic and establish its asymptotically normal under certain mild conditions. Simulation studies showed that our test procedure performs very well in many cases.

\vspace{0.2cm} \noindent{\bf Keywords}: Asymptotic normality; High-dimensional data; Large $p$, small $n$; $U$-statistics; Scale-invariant.
\end{abstract}
\section{Introduction}
In the past decades, high-dimensional data are increasingly encountered in statistical application from many areas, such as hyperspectral imagery, internet portals, microarray analysis and finance. A frequently encountered challenge in high-dimensional regression is the detection of relevant
variables. Identifying significant sets of genes which are associated with certain clinical outcome is very important in genomic studies, see Subramanian et al.
(2005), Efron and Tibshirani (2007) and Newton et al. (2007).  The main challenge of high-dimensional data is that the dimension $p$ is much larger than the sample sizes $n$. When this happens, many traditional statistical methods and theories may not necessarily work since they assume that $p$ keeps unchanged as $n$ increases. Recently, many efforts have been devoted to solve this problem. One is the variable selection method. Fan and Lv (2008) proposed the Sure Independence Screening (SIS) method based on a correlation learning to reduce the dimensionality from high to a moderate scale that is below sample size. Wang (2009) extended the classic Forward Regression method under an ultra-high dimensional setup.  The other method is hypothesis testing. To gain power and insight, it can be advantageous to look for influence not at the level of individual variables but rather
at the level of clusters of variables. Thus, A simultaneous test on linear regression coefficients in high-dimensional settings is needed. Goeman, Finos and van Houwelingen (2011) formulated an Empirical Bayes test via a score test on the hyper parameter of a prior distribution assumed on the regression coefficients. Zhong and Chen (2011) modified the classic $F$-statistic and proposed a $U$-statistic to examine the validity of the full model and extended their test to a linear model augmented with the factorial design setting.

However, both these two tests are not scalar invariant.  Intuitively speaking,
their test power would heavily depend on the underlying variance
magnitudes since they do not use the information from the diagonal elements of the
sample covariance, i.e., the variances of each variables. When all the components are (approximately) homogeneous
, they would be very powerful, whereas their superiority would be highly affected if the component variances differ much. In
practice, different components may have completely different physical or biological readings and thus certainly their scales
would not be the same. Hence, it is desirable to develop scalar-transformation-invariant test procedure which are able to integrate all the individual information in a relatively ``fair" way. In practice, due to confidentiality reasons, both the response and predictors will be firstly standardized to be zero mean and unit variance usually. When the dimension of predictors is low, the test efficiency is not impacted by this standardized procedure. However, when the dimension of predictors is ultra-high, there would be a large bias in the test procedure because the variance estimators are only root-$n$ consistent, see Feng et al. (2012) for the case in the high-dimensional two sample Behrens-Fisher problem. Thus, if we standardize the predictors firstly, Zhong and Chen (2011)'s test will not be reasonable when the dimension $p$ is ultra-high. This motivates us to discuss when the asymptotic normality of their test statistic still holds after standardizing the predictors. Thus, in this article, we proposed a novel test statistic which is scalar-invariant and provide the theoretical conditions when its asymptotic normality still holds. Simulation studies show that our proposed test has reasonable sizes and effective powers.

The remainder of the paper is organized as follows. In the next
section, we propose our test statistic and establish its asymptotic
normality. Simulation comparison is conducted in Section 3. All technical
details are provided in the Appendix.

\section{Test Statistics}

In this article, we consider the following linear regression model
\begin{align}
E(Y_i|\X_i)=\alpha+\X^{'}_i\bb, ~~\var(Y_i|\X_i)=\sigma^2
\end{align}
for $i=1,\cdots,n$ where $\X_1,\cdots, \X_n$ are independent and identically distributed $p$-dimensional covariates and $Y_1,\cdots,Y_n$ are independent responses, $\bb$ is the vector of regression coefficients, and $\alpha$ is a nuisance intercept. To make $\bb$ identifiable, we assume that $\bms=\var(\X_i)$ and $\R=\cor(\X_i)$ is positive definite. Our interest is in testing a high-dimensional hypothesis
\begin{align}\label{test}
H_0:\bb=\bb_0~~\rm{vs}~~H_1:\bb\not=\bb_0
\end{align}
for a specific $\bb_0 \in \mathbb{R}^p$. A classical method to deal with this problem is the famous $F$-test statistic
\begin{align*}
F_{n}=\frac{(\hat{\bb}-\bb_0)^{'}\A^{'}(\A(\U^{'}\U)^{-1}\A^{'})^{-1}(\hat{\bb}-\bb_0)/p}{\Y^{'}(\I_n-\U(\U^{'}\U)^{-1}\U^{'})\Y/(n-p-1)}
\end{align*}
where $\U=({\bf 1},\X)^{'}$, $A=(\bm 0, \I_p)$ and $\hat{\bb}$ is the least square estimator of $\bb$. Its advantages include: it is invariant under linear transformation, its exact distribution is known under the null hypothesis and it is powerful when the dimension of data is sufficiently small, compared with the sample sizes. However, Zhong and Chen (2011) showed that the power of $F$-test is adversely impacted by an increased dimension even $p<n-1$, reflecting a reduced degree of freedom in estimating $\sigma^2$ when the dimensionality is close to the sample size. Moreover, the $F$-test statistics is undefined when the dimension of data is greater than the within sample degrees of freedom since the pooled sample covariance matrices are not positive definite.
In order to overcome this issue, Goeman, Finos and van Houwelingen (2011) proposed an Empirical Bayes test, which is formulated via a score test on the hyper parameter of a prior distribution assumed on the regression coefficients. Their test statistics is
\begin{align}
G_n=\frac{(\Y-\hat{\alpha}-\X^{'}\bb_0)^{'}\X\X^{'}(\Y-\hat{\alpha}-\X^{'}\bb_0)}{n(\Y-\hat{\alpha}-\X^{'}\bb_0)^{'}(\Y-\hat{\alpha}-\X^{'}\bb_0)}
\end{align}
where $\hat{\alpha}$ is the sample mean of $Y$. The key feature of their method is to use Euclidian norm to replace the Mahalanobis norm since having $(\X^{'}\X)^{-1}$ is no longer beneficial when $p$ is larger than $n$. However, the power of $G_n$ is adversely impacted by $\bm \mu$, the mean of $\X$, which is a nuisance parameter in our interested test. Zhong and Chen (2011) consider a $U$-statistic
\begin{align}
Z_n=\frac{1}{4P_n^4}\sum^{*}(\X_{i_1}-\X_{i_2})^{'}(\X_{i_3}-\X_{i_4})(\Delta_{i_1}-\Delta_{i_2})(\Delta_{i_3}-\Delta_{i_4})
\end{align}
where $\Delta_i=Y_i-\X_i\bb_0$. Through this article, we use $\sum^{*}$ to denote summations over distinct indexes. For example, in $Z_n$, the summation is over the set $\{i_1\not=i_2\not=i_3\not=i_4\}$, for all $i_1,i_2,i_3,i_4\in\{1,\cdots,n\}$ and $P_n^m=\frac{n!}{(n-m)!}$. Obviously, $Z_n$ is not impacted by the nuisance parameter $\alpha$ and $\bm \mu$. They established the asymptotic normality of $Z_n$ under the diverging factor model ( Bai and Saranadasa 1996).

However, an obvious limitation of $G_n$ and $Z_n$ is that they are not invariant under scalar transformations. To this end, we standardize each component of $(\X_{i_1}-\X_{i_2})^{'}(\X_{i_3}-\X_{i_4})$ in $Z_n$ by the corresponding variance and propose a simple but effective test statistics,
\begin{align}
T_n=&\frac{1}{4P_n^4}\sum^{*}(\X_{i_1}-\X_{i_2})^{'}\D_S^{-1}(\X_{i_3}-\X_{i_4})(\Delta_{i_1}-\Delta_{i_2})(\Delta_{i_3}-\Delta_{i_4})
\end{align}
 where $\D_S$ is the diagonal matrix of pooled sample covariance matrix, that is
\[\D_S=\diag(\hat{\sigma}_1^2,\cdots,\hat{\sigma}_p^2)\]
where $\hat{\sigma}^2_{k}$ is the sample variance of $\{X_{ik}\}_{i=1}^n$, $k=1,\cdots,p$.
Obviously, $T_n$ is invariant to location shifts in both $\X_i$ and $Y_i$.  Thus, we assume, without loss of generality, that $\alpha=\mu=0$ in the rest of the article. Moreover, $T_n$ is invariant under the group of scalar transformations, say, $\X_{i}\to {\bf C}\X_i$ for $i=1,\cdots, n$ where ${\bf C}=\diag\{c_1,\cdots,c_p\}$ and $c_1,\cdots,c_p$ are non-zero constants.

In order to establish the asymptotic normality of $T_n$, we assume, like Bai and Saranadasa (1996), the following diverging factor model:
\begin{align*}
\X_i=\G \z_{i}+\bm \mu
\end{align*}
where $\G$ is a $p\times m$ matrix for some $m\ge p$ such that $\G \G^{'}=\bms$ and $\{\z_i\}_{i=1}^n$ are $m$-variate independent and identically distributed random vectors such that
\begin{align}
\label{chends}
\begin{array}{c}
 E(\z_i)=0, \ \var(\z_i)=\I_m,~E(z_{il}^{4})=3+\Delta,~ E(z_{il}^{8})=m_8\in (0,\infty),\\
  E(z_{ik_1}^{\alpha_1}z_{ik_2}^{\alpha_2}\cdots
  z_{ikq}^{\alpha_q})=E(z_{ik_1}^{\alpha_1})E(z_{ik_2}^{\alpha_2})\cdots
  E(z_{ikq}^{\alpha_q}),
 \end{array}
\end{align}
whenever $\sum_{k=1}^q\alpha_k\leq 8$ and $k_1\neq k_2\cdots\neq
k_q$. Additional, we need the following conditions to regulate for the `` large $p$, small $n$" is,
\begin{itemize}
\item[(C1)] $p(n)\to \infty$ as $n\to \infty$;
\item[(C2)] $\tr(\R^4)=o(\tr^2(\R^2))$;
\item[(C3)] $\frac{p^2}{n^2 \tr(\R^2)} \to 0$.
\end{itemize}
\remark 1 Both Condition (C1) and (C2) are similar to condition (2.8) in Zhong and Chen (2011). Since the estimator $\hat{\sigma}_k^2$ is only root-$n$ consistent, there would be a little bias term in the variance of  $T_n$. Fortunately, the bias term would be negligible when condition (C3) holds.
 To appreciate condition (C3), consider the simple case $\R=\I_p$, thus, the condition becomes $p=o(n^2)$. When $p$ gets larger, such as $p=O(n^2)$, the bias term in the variance of $T_n$ will no longer be negligible. Thus, we need a bias correction to solve this problem, see Feng et al. (2012) for more information.

In order to study the asymptotic power of our test, similar to Zhong and Chen (2011), we define the following local alternatives
\begin{align}\label{alter}
(\bb-\bb_0)^{'}\bms(\bb-\bb_0)&=o(1)\nonumber\\
(\bb-\bb_0)^{'}\bms\D^{-1}\bms\D^{-1}\bms(\bb-\bb_0)&=o(n^{-1}\tr(\R^2))
\end{align}
Note that the local alternatives (\ref{alter}) prescribe a smaller difference between $\bb$ and $\bb_0$. Similar to Zhong and Chen (2011), we also consider two different fixed alternatives which violate the first part of (\ref{alter}) in the Appendix. And we also demonstrate our proposed test can achieve at least $50\%$ power under these two fixed alternatives.

The following Theorem establishes the asymptotic normality of $T_n$ under the null or local alternative (\ref{alter}) hypothesis.
\begin{thm}
Assume conditions (C1)-(C3) hold, then under either $H_0$ or the local alternative (\ref{alter}), as $n\to\infty$,
\begin{align}
\frac{n}{\sigma^2\sqrt{2\tr(\R^2)}}\left(T_n-||\D^{-1/2}\bms(\bb-\bb_0)||^2\right) \cd N(0,1)
\end{align}
\end{thm}

To formulate a test procedure based on $T_n$, we need to estimate $\tr(\R^2)$ and $\sigma^2$ appeared in the asymptotic variance.
In order to reduce the computational work, we propose the following ratio consistent estimator of $\tr(\R^2)$,
\begin{align*}
\widehat{\tr(\R^2)}=\frac{1}{2P_n^4}\sum^{*} (\X_{i_1}-\X_{i_2})^{'}\D_S^{-1}(\X_{i_3}-\X_{i_4})(\X_{i_3}-\X_{i_2})^{'}\D_S^{-1}(\X_{i_1}-\X_{i_4})
\end{align*}
And the estimator of $\sigma^2$ under $H_0$ is
\begin{align*}
\hat{\sigma}^2=\frac{1}{n-1}\sum_{i=1}^n (Y_i-\X_i^{'}\bb_0-\bar{Y}+\bar{\X}^{'}\bb_0)^2
\end{align*}

\begin{pro}
Suppose the conditions in Theorem 1 hold. Then, as $n,p\to \infty$
\begin{align*}
\frac{\widehat{\tr(\R^2)}}{\tr(\R^2)}\cp 1
\end{align*}
\end{pro}
Apply Theorem 1 and the Slutsky Theorem, the proposed test rejects $H_0$ at a significant level $\alpha$ if
\begin{align}
nT_n\ge \sqrt{2\widehat{\tr(\R^2)}}\hat{\sigma}^2z_{\alpha}
\end{align}
where $z_{\alpha}$ is the upper-$\alpha$ quantile of $N(0,1)$.

Next, we discuss the power properties of the proposed test. According to Theorem 1, the power of our proposed test under the local alternative (\ref{alter}) is
\begin{align*}
\beta_{T_n}(||\bb-\bb_0||)=\Phi(-z_{\alpha}+\frac{n||\D^{-1/2}\bms(\bb-\bb_0)||^2}{\sqrt{2\tr(\R^2)}\sigma^2})
\end{align*}
where $\Phi$ is the standard normal distribution function. In comparison, Zhong and Chen (2011) show the power of their proposed test is
\begin{align*}
\beta_{Z_n}(||\bb-\bb_0||)=\Phi(-z_{\alpha}+\frac{n||\bms(\bb-\bb_0)||^2}{\sqrt{2\tr(\bms^2)}\sigma^2})
\end{align*}
Note that it is difficult to compare the proposed test with Zhong and Chen's (2011) test under general settings. Thus, in order to get a rough picture of the asymptotic power comparison between these two test, we consider the following representative cases:
\begin{itemize}
\item[(i)] The variances of all variables are equal to $\lambda$ and then $\bms=\lambda \R$. In this case,
\begin{align*}
\beta_{Z_n}(||\bb-\bb_0||)=\beta_{T_n}(||\bb-\bb_0||)=\Phi(-z_{\alpha}+\frac{n\lambda||\R(\bb-\bb_0)||^2}{\sqrt{2\tr(\R^2)}\sigma^2})
\end{align*}
\item[(ii)] $\bms(\bb-\bb_0)=\delta(1,1,\cdots,1)^{'}$. In this case,
\begin{align*}
\beta_{T_n}(||\bb-\bb_0||)&=\Phi(-z_{\alpha}+\frac{n\tr(\D^{-1})\delta^2}{\sqrt{2\tr(\R^2)}\sigma^2})\\
\beta_{Z_n}(||\bb-\bb_0||)&=\Phi(-z_{\alpha}+\frac{np\delta^2}{\sqrt{2\tr(\bms^2)}\sigma^2})
\end{align*}
According to the Cauchy inequality,
\[\tr^2(\D^{-1})\tr(\bms^2)\ge p^2 \tr(\R^2)\]
As a consequence,
\begin{align*}
\beta_{T_n}(||\bb-\bb_0||)\ge \beta_{Z_n}(||\bb-\bb_0||)
\end{align*}
When the variances of all the variables are equal, the two tests are equivalently powerful. Otherwise, the proposed test would be more preferable in this case.
\item[(iii)] $\bms$ is a diagonal matrix i.e. $\bms=\D$. The variances of the first half components are $\sigma_1^2$ and the rest are all $\sigma_2^2$. Assume $\beta_i-\beta_{0i}=\delta$, $i=1,\cdots,\lfloor\frac{p}{2}\rfloor$ and the others are all equal to zero. In this setting,
\begin{align*}
\beta_{T_n}(||\bb-\bb_0||)&=\Phi(-z_{\alpha}+\frac{n\sqrt{p}\sigma_1^2\delta^2}{2\sqrt{2}\sigma^2})\\
\beta_{Z_n}(||\bb-\bb_0||)&=\Phi(-z_{\alpha}+\frac{n\sqrt{p}\sigma_1^4\delta^2}{2\sqrt{\sigma_1^4+\sigma_2^4}\sigma^2})
\end{align*}
Thus, the asymptotic relative efficiency (ARE) of the porposed test
with respect to the Zhong and Chen's (2011) test would be
$\sqrt{\sigma_1^4+\sigma_2^4}/(\sqrt{2}\sigma_1^2)$. It is clear
that the proposed test is more powerful than Zhong and Chen's (2011) test if
$\sigma_1^2<\sigma_2^2$ and vice versa. This ARE has a positive
lower bound of $1/\sqrt{2}$ when $\sigma_1^2>>\sigma_2^2$, whereas
it can be arbitrarily large if $\sigma_1^2/\sigma_2^2$ is close to
zero.
\end{itemize}

\section{Simulation}
Here we report a simulation study designed to evaluate the performance of our proposed test (abbreviated as SF). For comparison purposes, we also conducted the test proposed by Zhong and Chen (2011) (abbreviated as ZC) and the Empirical Bayes test proposed by Goeman, Fino, and van Houwelingen (2011) (abbreviated as EB). We consider the following linear regression as Zhong and Chen (2011):
\begin{align}
Y_i=\X_i^{'}\bb+\varepsilon_i
\end{align}
and the hypotheses to be tested are
\begin{align}\label{test}
H_0:\bb=\bm 0_{p\times 1}~~\rm{vs}~~H_1:\bb\not=\bm 0_{p\times 1}
\end{align}
We consider two distributions for $\varepsilon_i$, one is $N(0,4)$, the other is centralized gamma distribution Gamma$(1,0.5)$. And $\X_i=(X_{i1},\cdots, X_{ip})$ are generated according to the following moving average model
\begin{align*}
X_{ij}=\rho_{1}Z_{ij}+\rho_{2}Z_{i(j+1)}+\cdots+\rho_{T} Z_{i(j+T-1)}+\mu_{ij}
\end{align*}
for $j=1,\cdots,p$ and $T<p$. Here $\{Z_{ij}\}_{j=1}^{p+T-1}$ are, respectively, i.i.d. random variables. We consider two scenarios for the innovation $\Z_{ij}$: (Scenario I) all the $\{Z_{ij}\}$ are from $N(0,1)$; (Scenario II) the first half components of $\{Z_{ij}\}_{j=1}^{p+T-1}$ are from $N(0,1)$, and the rest half components are from centralized Gamma$(4,1)$. The coefficient $\{\rho_{l}\}_{l=1}^{T}$ were generated independently from $U(0,1)$ and were kept fixe once generated through our simulations. And the means $\{\mu_i\}_{i=1}^p$ are also fixed constants generate from $U(2,3)$. We chose $T=10$ and $T=20$, to generate different covariances of $\X_i$. Similar to Zhong and Chen (2011), we consider two configurations of the alternative hypothesis $H_1$. One is ``nonsparse case'', which allocated first half of the $\bb$-components of equal magnitude to be nonzeros. The other is ``sparse case'', which has only the first five nonzero components of equal magnitude. In both case, we fixed $||\bb||^2$ at three levels: $0.03,0.06,0.09$. Here we only consider the case $p>n$ and chose $(n,p)=(30,100), (40,200), (50,400)$.

Table 1--2 and Table 3--4 reports the empirical sizes and powers with normally and centralized gamma distributed residuals, respectively. From Table 1 to Table 4, we observe that the empirical sizes are both reasonable for these three tests. And the the sizes of these two tests became closer to the nominal level 0.05 when $n$ and $p$ gets larger, which is similar to Zhong and Chen (2011)'s results. Moreover, from Table 1 and 3, we observe that when all the variances of components are equal (Scenario I), our proposed SF test performs similar to ZC tests and EB tests. Even though we need to estimate the variance of each component, our proposed SF test does not lose much information form the samples when the dimension $p$ is a small order of $n^2$. These findings are also consistent with the
asymptotic intuition in Section 2. However, when the variances of each components are not equal (Scenario II), our proposed SF test is clearly  much more powerful then the other two tests. This mainly due to the fact that ZC tests and EB tests are not scale-invariant. When the variances of variables are not equal, ZC tests and EB tests hardly capture the coefficient shifts with smaller variances and then it will be powerless in such cases. Thus, it is not strange that their performance are extremely poor in such cases.

   \begin{table}[ht]
          \centering
          \caption{Empirical size and power comparisons at 5\% significance for normal residual under Scenario I}
          \vspace{0.2cm}
     \renewcommand{\arraystretch}{1}
    \tabcolsep 9pt
        \begin{tabular}{ccccccccc}\hline
    &  & \multicolumn{3}{c}{T=10}& & \multicolumn{3}{c}{T=20}  \\ \cline{3-5} \cline{7-9}
   {\small $(n,p)$} &  {\small $||\bb||^2$}   &\multicolumn{1}{c}{SF} & \multicolumn{1}{c}{ZC} & \multicolumn{1}{c}{EB} & &  \multicolumn{1}{c}{SF} & \multicolumn{1}{c}{ZC} & \multicolumn{1}{c}{EB} \\\hline
   (a) nonsparse case \\
      (30,100)   &   0.00     &   0.06    &  0.06 & 0.05   &   &   0.06  &   0.06  & 0.06         \\
                 &   0.03     &   0.27    &  0.31 & 0.27   &   &   0.75  &   0.71  & 0.76          \\
                 &   0.06     &   0.53    &  0.54 & 0.46   &   &   0.93  &   0.91  & 0.95        \\
                 &   0.09     &   0.69    &  0.70 & 0.64   &   &   0.97  &   0.96  & 0.98         \\
      (40,200)   &   0.00     &   0.05    &  0.05 & 0.04   &   &   0.05  &   0.05  & 0.05         \\
                 &   0.03     &   0.25    &  0.25 & 0.37   &   &   0.78  &   0.77  & 0.82        \\
                 &   0.06     &   0.47    &  0.49 & 0.60   &   &   0.95  &   0.95  & 0.96     \\
                 &   0.09     &   0.63    &  0.69 & 0.76   &   &   0.98  &   0.97  & 1.00     \\
      (50,400)   &   0.00     &   0.05    &  0.05 & 0.06   &   &   0.05  &   0.05  & 0.04         \\
                 &   0.03     &   0.22    &  0.22 & 0.14   &   &   0.81  &   0.79  & 0.79      \\
                 &   0.06     &   0.45    &  0.42 & 0.33   &   &   0.94  &   0.95  & 1.00        \\
                 &   0.09     &   0.61    &  0.60 & 0.43   &   &   0.97  &   0.97  & 1.00       \\
   (b) sparse case \\
     (30,100)    &   0.03     &   0.12    &  0.13 & 0.07   &   &   0.15  &   0.15  & 0.16         \\
                 &   0.06     &   0.20    &  0.19 & 0.12   &   &   0.24  &   0.24  & 0.26        \\
                 &   0.09     &   0.24    &  0.23 & 0.15   &   &   0.31  &   0.30  & 0.35      \\
     (40,200)    &   0.03     &   0.11    &  0.12 & 0.14   &   &   0.16  &   0.14  & 0.17         \\
                 &   0.06     &   0.18    &  0.18 & 0.22   &   &   0.25  &   0.26  & 0.32     \\
                 &   0.09     &   0.24    &  0.24 & 0.32   &   &   0.40  &   0.39  & 0.48        \\
     (50,400)    &   0.03     &   0.10    &  0.10 & 0.11   &   &   0.12  &   0.12  & 0.15         \\
                 &   0.06     &   0.15    &  0.15 & 0.17   &   &   0.25  &   0.25  & 0.22         \\
                 &   0.09     &   0.18    &  0.21 & 0.28   &   &   0.34  &   0.33  & 0.33        \\\hline
                 \end{tabular}
             \end{table}

   \begin{table}[ht]
          \centering
          \caption{Empirical size and power comparisons at 5\% significance for normal residual under Scenario II}
          \vspace{0.2cm}
     \renewcommand{\arraystretch}{1}
    \tabcolsep 9pt
        \begin{tabular}{ccccccccc}\hline
    &  & \multicolumn{3}{c}{T=10}& & \multicolumn{3}{c}{T=20}  \\ \cline{3-5} \cline{7-9}
   {\small $(n,p)$} &  {\small $||\bb||^2$}   &\multicolumn{1}{c}{SF} & \multicolumn{1}{c}{ZC} & \multicolumn{1}{c}{EB} & &  \multicolumn{1}{c}{SF} & \multicolumn{1}{c}{ZC} & \multicolumn{1}{c}{EB} \\\hline
   (a) nonsparse case \\
      (30,100)   &   0.00    &   0.06      &   0.06 & 0.06  &    & 0.05   & 0.07 & 0.06    \\
                 &   0.03    &   0.29      &   0.10 & 0.12  &    & 0.70   & 0.28 & 0.34   \\
                 &   0.06    &   0.48      &   0.17 & 0.18  &    & 0.92   & 0.50 & 0.53   \\
                 &   0.09    &   0.63      &   0.23 & 0.27  &    & 0.96   & 0.59 & 0.59   \\
      (40,200)   &   0.00    &   0.05      &   0.05 & 0.06  &    & 0.05   & 0.05 & 0.04  \\
                 &   0.03    &   0.35      &   0.15 & 0.10  &    & 0.77   & 0.31 & 0.32   \\
                 &   0.06    &   0.55      &   0.19 & 0.12  &    & 0.95   & 0.48 & 0.51   \\
                 &   0.09    &   0.65      &   0.23 & 0.14  &    & 0.97   & 0.59 & 0.63  \\
      (50,400)   &   0.00    &   0.05      &   0.05 & 0.05  &    & 0.06   & 0.06 & 0.05   \\
                 &   0.03    &   0.23      &   0.10 & 0.07  &    & 0.80   & 0.34 & 0.33  \\
                 &   0.06    &   0.45      &   0.15 & 0.10  &    & 0.95   & 0.48 & 0.49  \\
                 &   0.09    &   0.60      &   0.17 & 0.14  &    & 0.99   & 0.55 & 0.60 \\
   (b) sparse case \\
     (30,100)    &   0.03    &   0.09      &   0.06 & 0.05  &    & 0.16   & 0.07 & 0.06 \\
                 &   0.06    &   0.18      &   0.07 & 0.07  &    & 0.28   & 0.10 & 0.11    \\
                 &   0.09    &   0.23      &   0.09 & 0.07  &    & 0.40   & 0.11 & 0.12   \\
     (40,200)    &   0.03    &   0.13      &   0.13 & 0.07  &    & 0.16   & 0.09 & 0.08   \\
                 &   0.06    &   0.16      &   0.14 & 0.09  &    & 0.28   & 0.11 & 0.10   \\
                 &   0.09    &   0.24      &   0.14 & 0.11  &    & 0.35   & 0.13 & 0.12   \\
     (50,400)    &   0.03    &   0.07      &   0.05 & 0.07  &    & 0.15   & 0.11 & 0.06   \\
                 &   0.06    &   0.14      &   0.05 & 0.09  &    & 0.25   & 0.12 & 0.09   \\
                 &   0.09    &   0.18      &   0.06 & 0.09  &    & 0.33   & 0.14 & 0.12  \\\hline
                 \end{tabular}
             \end{table}

 \begin{table}[ht]
          \centering
          \caption{Empirical size and power comparisons at 5\% significance for centralized gamma residual under Scenario I}
          \vspace{0.2cm}
     \renewcommand{\arraystretch}{1}
    \tabcolsep 9pt
        \begin{tabular}{ccccccccc}\hline
    &  & \multicolumn{3}{c}{T=10}& & \multicolumn{3}{c}{T=20}  \\ \cline{3-5} \cline{7-9}
   {\small $(n,p)$} &  {\small $||\bb||^2$}   &\multicolumn{1}{c}{SF} & \multicolumn{1}{c}{ZC} & \multicolumn{1}{c}{EB} & &  \multicolumn{1}{c}{SF} & \multicolumn{1}{c}{ZC} & \multicolumn{1}{c}{EB} \\\hline
   (a) nonsparse case \\
      (30,100)   &   0.00     &   0.05    &  0.06  & 0.06  &  &   0.06  &   0.06          & 0.06          \\
                 &   0.03     &   0.33    &  0.31  & 0.25  &  &   0.79  &   0.76          & 0.84            \\
                 &   0.06     &   0.56    &  0.58  & 0.43  &  &   0.90  &   0.88          & 0.95            \\
                 &   0.09     &   0.70    &  0.72  & 0.58  &  &   0.96  &   0.95          & 0.98           \\
      (40,200)   &   0.00     &   0.05    &  0.06  & 0.06  &  &   0.06  &   0.06          & 0.03            \\
                 &   0.03     &   0.29    &  0.31  & 0.30  &  &   0.83  &   0.82          & 0.96           \\
                 &   0.06     &   0.48    &  0.48  & 0.53  &  &   0.95  &   0.94          & 0.99        \\
                 &   0.09     &   0.65    &  0.64  & 0.71  &  &   0.98  &   0.98          & 1.00              \\
      (50,400)   &   0.00     &   0.05    &  0.05  & 0.07  &  &   0.05  &   0.05          & 0.06            \\
                 &   0.03     &   0.29    &  0.28  & 0.39  &  &   0.80  &   0.80          & 0.85           \\
                 &   0.06     &   0.53    &  0.53  & 0.63  &  &   0.95  &   0.94          & 0.98          \\
                 &   0.09     &   0.66    &  0.65  & 0.78  &  &   0.98  &   0.98          & 0.99         \\
   (b) sparse case \\
     (30,100)    &   0.03     &   0.11    &  0.14  & 0.15  &  &   0.20  &   0.20          & 0.17          \\
                 &   0.06     &   0.18    &  0.19  & 0.22  &  &   0.32  &   0.32          & 0.30          \\
                 &   0.09     &   0.25    &  0.27  & 0.32  &  &   0.40  &   0.42          & 0.41          \\
     (40,200)    &   0.03     &   0.12    &  0.12  & 0.10  &  &   0.21  &   0.21          & 0.26       \\
                 &   0.06     &   0.19    &  0.18  & 0.17  &  &   0.32  &   0.31          & 0.46       \\
                 &   0.09     &   0.24    &  0.23  & 0.19  &  &   0.39  &   0.39          & 0.57       \\
     (50,400)    &   0.03     &   0.11    &  0.10  & 0.10  &  &   0.15  &   0.14          & 0.14       \\
                 &   0.06     &   0.15    &  0.15  & 0.20  &  &   0.25  &   0.23          & 0.28       \\
                 &   0.09     &   0.21    &  0.20  & 0.24  &  &   0.36  &   0.35          & 0.41      \\\hline
                 \end{tabular}
             \end{table}

 \begin{table}[ht]
          \centering
          \caption{Empirical size and power comparisons at 5\% significance for centralized gamma residual under Scenario II}
          \vspace{0.2cm}
   \renewcommand{\arraystretch}{1}
    \tabcolsep 9pt
        \begin{tabular}{ccccccccc}\hline
    &  & \multicolumn{3}{c}{T=10}& & \multicolumn{3}{c}{T=20}  \\ \cline{3-5} \cline{7-9}
   {\small $(n,p)$} &  {\small $||\bb||^2$}   &\multicolumn{1}{c}{SF} & \multicolumn{1}{c}{ZC} & \multicolumn{1}{c}{EB} & &  \multicolumn{1}{c}{SF} & \multicolumn{1}{c}{ZC} & \multicolumn{1}{c}{EB} \\\hline
   (a) nonsparse case \\
      (30,100)   &   0.00   &   0.06      &   0.06  & 0.05 &  & 0.06   & 0.06  & 0.05  \\
                 &   0.03   &   0.38      &   0.13  & 0.12 &  & 0.76   & 0.38  & 0.35  \\
                 &   0.06   &   0.56      &   0.17  & 0.17 &  & 0.93   & 0.57  & 0.55  \\
                 &   0.09   &   0.70      &   0.25  & 0.21 &  & 0.98   & 0.66  & 0.66  \\
      (40,200)   &   0.00   &   0.06      &   0.05  & 0.05 &  & 0.05   & 0.06  & 0.05  \\
                 &   0.03   &   0.30      &   0.11  & 0.12 &  & 0.75   & 0.28  & 0.25  \\
                 &   0.06   &   0.49      &   0.18  & 0.18 &  & 0.97   & 0.47  & 0.38  \\
                 &   0.09   &   0.63      &   0.23  & 0.23 &  & 0.99   & 0.61  & 0.50  \\
      (50,400)   &   0.00   &   0.06      &   0.06  & 0.07 &  & 0.06   & 0.06  & 0.05  \\
                 &   0.03   &   0.30      &   0.12  & 0.10 &  & 0.79   & 0.34  & 0.46   \\
                 &   0.06   &   0.49      &   0.16  & 0.14 &  & 0.96   & 0.45  & 0.64 \\
                 &   0.09   &   0.62      &   0.19  & 0.17 &  & 0.99   & 0.53  & 0.71 \\
   (b) sparse case \\
     (30,100)    &   0.03   &   0.10      &   0.09  & 0.06 &  & 0.19   & 0.09  & 0.08 \\
                 &   0.06   &   0.20      &   0.09  & 0.08 &  & 0.30   & 0.10  & 0.10  \\
                 &   0.09   &   0.27      &   0.09  & 0.08 &  & 0.40   & 0.14  & 0.12  \\
     (40,200)    &   0.03   &   0.12      &   0.10  & 0.06 &  & 0.17   & 0.08  & 0.06  \\
                 &   0.06   &   0.20      &   0.10  & 0.08 &  & 0.32   & 0.10  & 0.09 \\
                 &   0.09   &   0.25      &   0.11  & 0.09 &  & 0.43   & 0.13  & 0.10  \\
     (50,400)    &   0.03   &   0.16      &   0.08  & 0.05 &  & 0.15   & 0.11  & 0.07 \\
                 &   0.06   &   0.22      &   0.10  & 0.07 &  & 0.25   & 0.10  & 0.09  \\
                 &   0.09   &   0.28      &   0.10  & 0.08 &  & 0.37   & 0.12  & 0.11  \\\hline
                 \end{tabular}
             \end{table}

\section{Appendix}
\subsection{Proof of Theorem 1}
Define $D$ the diagonal matrix of covariance matrix, that is
\begin{align*}
D=\diag(\sigma_1^2,\cdots,\sigma_p^2).
\end{align*}
Thus, we can rewrite $T_n$ as follow
\begin{align*}
T_n=&\frac{1}{4P_n^4}\sum^{*}(\X_{i_1}-\X_{i_2})^{'}\D^{-1}(\X_{i_3}-\X_{i_4})(\Delta_{i_1}-\Delta_{i_2})(\Delta_{i_3}-\Delta_{i_4})\\
&+\frac{1}{4P_n^4}\sum^{*}(\X_{i_1}-\X_{i_2})^{'}(\D_S^{-1}-\D^{-1})(\X_{i_3}-\X_{i_4})(\Delta_{i_1}-\Delta_{i_2})(\Delta_{i_3}-\Delta_{i_4})\\
\doteq &T_{n1}+T_{n2}
\end{align*}
Define
\begin{align*}
\phi(i_1,i_2,i_3,i_4)=\frac{1}{4}(\X_{i_1}-\X_{i_2})^{'}\D^{-1}(\X_{i_3}-\X_{i_4})(\Delta_{i_1}-\Delta_{i_2})(\Delta_{i_3}-\Delta_{i_4})
\end{align*}
And then we symmetrize $\phi$ by
\begin{align*}
h(W_i,W_j,W_k,W_l)=\frac{1}{3}\{\phi(i,j,k,l)+\phi(i,k,j,l)+\phi(i,l,j,k)\}
\end{align*}
where $W_i=(\X_i^{'},\varepsilon_i)^{'}$ and $\varepsilon_i=Y_i-\X_i^{'}\bb$. Thus
\begin{align*}
T_{n1}=\frac{1}{C_n^4}\sum_{C_{n,4}} h(W_i,W_j,W_k,W_l).
\end{align*}
Define $\delta_{\bb}=\bb-\bb_0$. After some tedious calculation, we can obtain the projections of $h$ are, respectively,
\begin{align*}
h_1(W_1)=&\frac{1}{2}\delta_{\bb}^{'}(\X_1\X_1^{'}+\bms)\D^{-1}\bms \delta_{\bb}+\frac{1}{2}\varepsilon_1\X_1^{'}\D^{-1}\bms\delta_{\bb}\\
h_2(W_1,W_2)=&\frac{1}{6}\Big\{\delta_{\bb}^{'}(\X_1-\X_2)(\X_1-\X_2)^{'}\D^{-1}\bms\delta_{\bb}+(\varepsilon_1-\varepsilon_2)(\X_1-\X_2)^{'}\D^{-1}\bms \delta_{\bb}\\
&+\left(\delta_{\bb}^{'}(\X_1\X_1^{'}+\bms)+\varepsilon_1\X_1^{'}\right)\D^{-1} \left((\X_2\X_2^{'}+\bms)\delta_{\bb}+\varepsilon_2\X_2\right)\Big\}\\
h_3(W_1,W_2,W_3)=&\frac{1}{12}\left((\X_1-\X_2)^{'}\delta_{\bb}+(\varepsilon_1-\varepsilon_2)\right)\D^{-1}(\X_1-\X_2)^{'} \left((\X_3\X_3^{'}+\bms)\delta_{\bb}+\varepsilon_3\X_3\right)\\
&+\frac{1}{12}\left((\X_1-\X_3)^{'}\delta_{\bb}+(\varepsilon_1-\varepsilon_3)\right)\D^{-1}(\X_1-\X_3)^{'} \left((\X_2\X_2^{'}+\bms)\delta_{\bb}+\varepsilon_2\X_2\right)\\
&+\frac{1}{12}\left((\X_2-\X_3)^{'}\delta_{\bb}+(\varepsilon_2-\varepsilon_3)\right)\D^{-1}(\X_2-\X_3)^{'} \left((\X_1\X_1^{'}+\bms)\delta_{\bb}+\varepsilon_1\X_1\right)
\end{align*}
Define $B_1=\delta_{\bb}^{'}\bms \delta_{\bb}$, $B_2=\delta_{\bb}^{'}\bms\D^{-1}\bms \delta_{\bb}$, $B_3=\delta_{\bb}^{'}\bms \D^{-1}\bms \D^{-1}\bms \delta_{\bb}$ and $A_0=\G^{'}\G$, $A_1=\G^{'}\delta_{\bb}\delta_{\bb}^{'}\G$, $A_2=\G^{'}\D\bms \delta_{\bb}\delta_{\bb}^{'}\bms \D^{-1}\G$, $A_3=\G^{'}\R\G$. Then,
\begin{align*}
\var(h_1)=&\frac{1}{4}\left\{(B_1+\sigma^2)B_3+B_2^2+\Delta\tr(A_1\circ A_2)\right\}\\
\var(h_2)=&\frac{1}{36}\Big\{\sigma^4\tr(\R^2)+21B_2^2+22B_1B_3+22\sigma^2B_3+B_1^2\tr(\R^2)+2\sigma^2\tr(\R^2)B_1\\
&+2\Delta(B_1+\sigma^2)\tr(A_1\circ A_3)+20\Delta\tr(A_1\circ A_2)+\Delta^2\tr[(A_0\diag(A_1))^2]\Big\}\\
\var(h)=&\frac{1}{24}\Big\{12\sigma^4\tr(\R^2)+45B_2^2+65B_1B_3+40\sigma^2B_3+10B_1^2\tr(\R^2)+24\sigma^2\tr(\R^2)B_1\\
&+12\Delta(B_1+\sigma^2)\tr(A_1\circ A_3)+37\Delta\tr(A_1\circ A_2)+4\Delta^2\tr[(A_0\diag(A_1))^2]\Big\}
\end{align*}
Thus, $\var(h_2)$ and $\var(h)$ are of the same order. Next, taking the same procedure as Zhong and Chen (2011), under the condition (\ref{alter}), we can show that
\begin{align*}
T_{n1}=||\D^{-1/2}\bms(\bb-\bb_0)||^2+\frac{2}{n(n-1)}\sum_{i<j}\varepsilon_i\varepsilon_j\X_i^{'}\X_j+o_p(\sqrt{\var(T_{n1})})
\end{align*}
And then, similar to Zhong and Chen (2011),  we can easily obtain that
\begin{align}
\frac{n}{\sigma^2\sqrt{2\tr(\R^2)}}\left(T_{n1}-||\D^{-1/2}\bms(\bb-\bb_0)||^2\right) \cd N(0,1)
\end{align}
by applying the martingale central limit theorem (Hall and Heyde 1980).

In order to proof Theorem 1, we only need to show that $T_{n_2}=o(\frac{1}{n}\sigma^2\sqrt{2\tr(\R^2)})$.
\begin{align*}
T_{n2}=&\frac{1}{4P_n^4}\sum^*\sum_{k=1}^p(x_{i_1k}-x_{i_2k})(x_{i_3k}-x_{i_4k})(\Delta_{i_1}-\Delta_{i_2})(\Delta_{i_3}-\Delta_{i_4})(\hat{\sigma}_k^{-2}-\sigma_k^{-2})\\
=&\frac{1}{4P_n^4}\sum^*\sum_{k=1}^p(x_{i_1k}-x_{i_2k})(x_{i_3k}-x_{i_4k})(\Delta_{i_1}-\Delta_{i_2})(\Delta_{i_3}-\Delta_{i_4})(\hat{\sigma}_k^{-2}-\sigma_k^{-2})\\
=&\frac{1}{4P_n^4}\sum^*\sum_{k=1}^p(x_{i_1k}-x_{i_2k})(x_{i_3k}-x_{i_4k})(\Delta_{i_1}-\Delta_{i_2})(\Delta_{i_3}-\Delta_{i_4})(\sigma_k^{2}-\hat{\sigma}_k^{2})\sigma_k^{-4}\\
&+\frac{1}{4P_n^4}\sum^*\sum_{k=1}^p(x_{i_1k}-x_{i_2k})(x_{i_3k}-x_{i_4k})(\Delta_{i_1}-\Delta_{i_2})(\Delta_{i_3}-\Delta_{i_4})(1-\hat{\sigma}_k^2\sigma_k^{-2})^2\hat{\sigma}_k^{-2}\\
\doteq & A_1+A_2
\end{align*}
Firstly, we will show that $E(A_1^2)=o(\frac{1}{n^2}\sigma^4\tr(\R^2))$.
\begin{align*}
&E(A_1^2)\\
=&\frac{1}{16(P_n^4)^2}\sum_{k=1}^p\sum_{l=1}^p E\Bigg\{ \sum^*_{i_1,i_2,i_3,i_4}(x_{i_1k}-x_{i_2k})(x_{i_3k}-x_{i_4k})(\Delta_{i_1}-\Delta_{i_2})(\Delta_{i_3}-\Delta_{i_4})(\sigma_k^{2}-\hat{\sigma}_k^{2})\sigma_k^{-4}\\
&\times \sum^*_{i_5,i_6,i_7,i_8}(x_{i_5l}-x_{i_6l})(x_{i_7l}-x_{i_8l})(\Delta_{i_5}-\Delta_{i_6})(\Delta_{i_7}-\Delta_{i_8})(\sigma_l^{2}-\hat{\sigma}_l^{2})\sigma_l^{-4}\Bigg\}\\
=&\frac{1}{16(P_n^4)^2}\sum_{k=1}^p\sum_{l=1}^p E\Bigg\{ \sum^*_{i_1,i_2,i_3,i_4,i_5,i_6}(\Delta_{i_1}-\Delta_{i_2})(\Delta_{i_3}-\Delta_{i_4})(\Delta_{i_1}-\Delta_{i_5})(\Delta_{i_3}-\Delta_{i_6})\\
&\times\sigma_l^{-4}\sigma_k^{-4} (x_{i_1k}-x_{i_2k})(x_{i_3k}-x_{i_4k})(x_{i_1l}-x_{i_5l})(x_{i_3l}-x_{i_6l})\\
&\times \left(\frac{1}{n}\sum_{i=1}^n(\sigma_k^2- x_{ik}^2)+\frac{2}{n(n-1)}\sum_{i\not=j}x_{ik}x_{jk}\right)\left(\frac{1}{n}\sum_{i=1}^n(\sigma_l^2- x_{il}^2)+\frac{2}{n(n-1)}\sum_{i\not=j}x_{il}x_{jl}\right)\Bigg\}\\
=&\frac{1}{16n^2(P_n^4)^2}\sum_{k=1}^p\sum_{l=1}^p E\Bigg\{ \sum^*_{i_1,i_2,i_3,i_4,i_5,i_6}\sum_{i=1}^n\sum_{j=1}^n(\Delta_{i_1}-\Delta_{i_2})(\Delta_{i_3}-\Delta_{i_4})(\Delta_{i_1}-\Delta_{i_5})(\Delta_{i_3}-\Delta_{i_6})\\
&\times\sigma_l^{-4}\sigma_k^{-4} (x_{i_1k}-x_{i_2k})(x_{i_3k}-x_{i_4k})(x_{i_1l}-x_{i_5l})(x_{i_3l}-x_{i_6l})(\sigma_k^2- x_{ik}^2)(\sigma_l^2- x_{jl}^2)\Bigg\}\\
&+\frac{1}{16n^2(n-1)(P_n^4)^2}\sum_{k=1}^p\sum_{l=1}^p E\Bigg\{ \sum^*_{i_1,i_2,i_3,i_4,i_5,i_6}\sum_{i_7=1}^n\sum_{i_8\not=i_9}^n(\Delta_{i_1}-\Delta_{i_2})(\Delta_{i_3}-\Delta_{i_4})(\Delta_{i_1}-\Delta_{i_5})\\
&\times(\Delta_{i_3}-\Delta_{i_6})\sigma_l^{-4}\sigma_k^{-4} (x_{i_1k}-x_{i_2k})(x_{i_3k}-x_{i_4k})(x_{i_1l}-x_{i_5l})(x_{i_3l}-x_{i_6l})(\sigma_k^2- x_{i_7k}^2)x_{i_8l}x_{i_9l}\Bigg\}\\
&+\frac{1}{16n^2(n-1)(P_n^4)^2}\sum_{k=1}^p\sum_{l=1}^p E\Bigg\{ \sum^*_{i_1,i_2,i_3,i_4,i_5,i_6}\sum_{i_7=1}^n\sum_{i_8\not=i_9}^n(\Delta_{i_1}-\Delta_{i_2})(\Delta_{i_3}-\Delta_{i_4})(\Delta_{i_1}-\Delta_{i_5})\\
&\times(\Delta_{i_3}-\Delta_{i_6})\sigma_l^{-4}\sigma_k^{-4} (x_{i_1k}-x_{i_2k})(x_{i_3k}-x_{i_4k})(x_{i_1l}-x_{i_5l})(x_{i_3l}-x_{i_6l})(\sigma_l^2- x_{i_7l}^2)x_{i_8k}x_{i_9k}\Bigg\}\\
&+\frac{1}{16n^2(n-1)^2(P_n^4)^2}\sum_{k=1}^p\sum_{l=1}^p E\Bigg\{ \sum^*_{i_1,i_2,i_3,i_4,i_5,i_6}\sum_{i_7\not=i_8}^n\sum_{i_9\not=i_{10}}^n(\Delta_{i_1}-\Delta_{i_2})(\Delta_{i_3}-\Delta_{i_4})(\Delta_{i_1}-\Delta_{i_5})\\
&\times(\Delta_{i_3}-\Delta_{i_6})\sigma_l^{-4}\sigma_k^{-4} (x_{i_1k}-x_{i_2k})(x_{i_3k}-x_{i_4k})(x_{i_1l}-x_{i_5l})(x_{i_3l}-x_{i_6l})x_{i_7k}x_{i_8k}x_{i_{9}l}x_{i_{10}l}\Bigg\}\\
\doteq & A_{11}+A_{12}+A_{13}+A_{14}
\end{align*}
After some tedious calculation, under condition (\ref{alter}), we can obtain that
\begin{align*}
A_{11}=&O(\frac{p^2}{n^4}) \left(\sigma_l^{-4}\sigma_k^{-4}(a_{kl}\sigma_k^2-E(x_{ik}^3x_{il}))(a_{kl}\sigma_l^2-E(x_{ik}x_{il}^3)\right)\\
&+O(\frac{p^2}{n^4})\left(\sigma_l^{-4}\sigma_k^{-4}a_{kl}(a_{kl}\sigma_k^2\sigma_l^2-\sigma_l^2E(x_{ik}^3x_{il})-\sigma_k^2E(x_{ik}x_{il}^3)+E(x_{ik}^3x_{il}^3))\right)+o(\frac{1}{n^2}\sigma^4\tr(\R^2))
\end{align*}
where $\bms=(a_{ij})_{i,j=1,\cdots,p}$. Define $\G=(v_{ij})$, according to the multivariate model, we can show that
\begin{align*}
E(x_{ik}^3x_{il})=&E\left(\left(\sum_{i=1}^m v_{ki}z_i\right)^3\left(\sum_{j=1}^m v_{lj}z_j\right)\right)=E\left(\sum_{i=1}^m\sum_{j=1}^m\sum_{s=1}^m\sum_{t=1}^m v_{ki}v_{kj}v_{ks}v_{lt}z_iz_jz_sz_t\right)\\
=&(3+\Delta)\sum_{i=1}^m v_{ki}^3v_{li}+3\sum_{i\not=j}^m v_{ki}^2v_{kj}v_{lj}=\Delta\sum_{i=1}^m v_{ki}^3v_{li}+3\sigma_{k}^2a_{kl}\\
\le & \Delta \sqrt{\sum_{i=1}^m v_{ki}^4 \sum_{i=1}^m v_{ki}^2v_{li}^2}+3\sigma_{k}^2a_{kl}
\le  \Delta \sqrt{\left(\sum_{i=1}^m v_{ki}^2\right)^3\left(\sum_{i=1}^m v_{li}^2\right)}+3\sigma_{k}^2a_{kl}\\
=& \Delta \sigma_{k}^3\sigma_{l}+3\sigma_{k}^2a_{kl}
\end{align*}
Define $E(z_i^6)=\Psi<+\infty$,
\begin{align*}
E(x_{il}^3x_{ik}^3)=&E\left(\left(\sum_{i=1}^m v_{ki}z_i\right)^3\left(\sum_{j=1}^m v_{lj}z_j\right)^3\right)\\
=&E\left(\sum_{i=1}^m\sum_{j=1}^m\sum_{s=1}^m\sum_{t=1}^m\sum_{r=1}^m\sum_{w=1}^m v_{ki}v_{kj}v_{kr}v_{ls}v_{lt}v_{wl}z_iz_jz_rz_wz_sz_t\right)\\
=&\sum_{i=1}^m\sum_{j=1}^m\sum_{s=1}^m\sum_{t=1}^m\sum_{r=1}^m\sum_{w=1}^m v_{ki}v_{kj}v_{kr}v_{ls}v_{lt}v_{wl}E(z_iz_jz_rz_wz_sz_t)\\
=&\Psi\sum_{i=1}^m v_{ki}^3v_{li}^3 +(27+9\Delta)\sum_{i\not=j}^m v_{ki}^2v_{kj}v_{li}^2v_{lj}+(27+9\Delta)\sum_{i\not=j}^m v_{ki}^3v_{li}v_{lj}^2+9\sum_{i\not=j\not=s}^m v_{ki}^2v_{lj}^2v_{ks}v_{ls}\\
\le &\frac{\Psi}{2}\sum_{i=1}^m v_{ki}^2v_{li}^2(v_{ki}^2+v_{li}^2)+\frac{27+9\Delta}{2}\sum_{i=1}^m\sum_{j=1}^mv_{ki}^2v_{li}^2(v_{kj}^2+v_{lj}^2)\\
&+\frac{27+9\Delta}{2}\sum_{i=1}^m\sum_{j=1}^mv_{ki}^2v_{lj}^2(v_{ki}^2+v_{li}^2)+\frac{9}{2}\sum_{i=1}^m \sum_{j=1}^m\sum_{s=1}^m v_{ki}^2v_{lj}^2(v_{ks}^2+v_{ls}^2)\\
\le &\frac{1}{2}(\Psi+63+18\Delta)(\sigma_{k}^4\sigma_{l}^2+\sigma_{l}^4 \sigma_{k}^2)
\end{align*}
Thus, we obtain that $A_{11}=O(\frac{p^2}{n^4})+o(\frac{1}{n^2}\sigma^4\tr(\R^2))=o(\frac{1}{n^2}\sigma^4\tr(\R^2))$ by the condition (C3). Taking the same procedure as $A_{11}$, we can show that $A_{12}, A_{13}, A_{14}$ are all $\frac{1}{n^2}\sigma^4\tr(\R^2)$. Here, we obtain the result that $E(A_1^2)=o(\frac{1}{n^2}\sigma^4\tr(\R^2))$.

Next, we rewrite $A_2$ as follows,
\begin{align*}
A_2=&\sum_{k=1}^p\left(\frac{1}{4P_n^4}\sum^*(x_{i_1k}-x_{i_2k})(x_{i_3k}-x_{i_4k})(\Delta_{i_1}-\Delta_{i_2})(\Delta_{i_3}-\Delta_{i_4})\hat{\sigma}_k^{-2}\right)(1-\hat{\sigma}_k^2\sigma_k^{-2})^2\\
&\doteq \sum_{k=1}^p C_kD_k
\end{align*}
By the Cauchy inequality, we obtain that
\begin{align*}
E(A_2^2)=&E\left(\left(\sum_{k=1}^p C_kD_k\right)^2\right)\le E\left(\left(\sum_{k=1}^p C_k^2\right)\left(\sum_{k=1}^p D_k^2\right)\right)\\
&\le\sqrt{E\left(\left(\sum_{k=1}^p C_k^2\right)^2\right)E\left(\left(\sum_{k=1}^p D_k^2\right)^2\right)}
\end{align*}
%\begin{align*}
%E\left(\left(\sum_{k=1}^p C_k^2\right)^2\right)=E\left(\sum_{k=1}^p\sum_{l=1}^p C_k^2C_l^2\right)
%\end{align*}
Taking the same procedure as $A_{11}$, we can show that $E(C_k^2C_l^2)=O(n^{-4})$ and $E(D_k^2D_l^2)=O(n^{-4})$. Thus, $E(A_2^2)=O(\frac{p^2}{n^{4}})=o(\frac{1}{n^2}\sigma^4\tr(\R^2))$ by the condition (C3). Here we proof the results.

\subsection{Proof of Proposition 1}
Firstly, after some tedious calculation, we can rewrite $\widehat{\tr(\R^2)}$ as follow,
\begin{align*}
&\frac{1}{2P_n^4}\sum^{*} (\X_{i_1}-\X_{i_2})^{'}\D_S^{-1}(\X_{i_3}-\X_{i_4})(\X_{i_3}-\X_{i_2})^{'}\D_S^{-1}(\X_{i_1}-\X_{i_4})\\
=&\frac{1}{P_n^2}\sum^{*}(\X_{i_1}^{'}\D_S^{-1}\X_{i_2})^2-\frac{2}{P_n^3}\sum^{*}\X_{i_1}^{'}\D_S^{-1}\X_{i_2}\X_{i_2}^{'}\D_S^{-1}\X_{i_3}
+\frac{1}{P_n^4}\sum^{*}\X_{i_1}^{'}\D_S^{-1}\X_{i_2}\X_{i_3}^{'}\D_S^{-1}\X_{i_4}
\end{align*}
Taking the same procedure as Theorem 1, we can show that
\begin{align*}
\frac{1}{P_n^2}\sum^{*}(\X_{i_1}^{'}\D_S^{-1}\X_{i_2})^2&=\frac{1}{P_n^2}\sum^{*}(\X_{i_1}^{'}\D^{-1}\X_{i_2})^2+o_p(\tr(\R^2))\\
\frac{2}{P_n^3}\sum^{*}\X_{i_1}^{'}\D_S^{-1}\X_{i_2}\X_{i_2}^{'}\D_S^{-1}\X_{i_3}&=\frac{2}{P_n^3}\sum^{*}\X_{i_1}^{'}\D^{-1}\X_{i_2}\X_{i_2}^{'}\D^{-1}\X_{i_3}+o_p(\tr(\R^2))\\
\frac{1}{P_n^4}\sum^{*}\X_{i_1}^{'}\D_S^{-1}\X_{i_2}\X_{i_3}^{'}\D_S^{-1}\X_{i_4}&=\frac{1}{P_n^4}\sum^{*}\X_{i_1}^{'}\D^{-1}\X_{i_2}\X_{i_3}^{'}\D^{-1}\X_{i_4}+o_p(\tr(\R^2))
\end{align*}
Thus,
\begin{align*}
\widehat{\tr(\R^2)}=&\frac{1}{P_n^2}\sum^{*}(\X_{i_1}^{'}\D^{-1}\X_{i_2})^2-\frac{2}{P_n^3}\sum^{*}\X_{i_1}^{'}\D^{-1}\X_{i_2}\X_{i_2}^{'}\D^{-1}\X_{i_3}\\
&+\frac{1}{P_n^4}\sum^{*}\X_{i_1}^{'}\D^{-1}\X_{i_2}\X_{i_3}^{'}\D^{-1}\X_{i_4}+o_p(\tr(\R^2))\\
=&\frac{1}{P_n^2}\sum^{*}(\tilde{\X}_{i_1}^{'}\tilde{\X}_{i_2})^2-\frac{2}{P_n^3}\sum^{*}\tilde{\X}_{i_1}^{'}\tilde{\X}_{i_2}\tilde{\X}_{i_2}^{'}\tilde{\X}_{i_3}
+\frac{1}{P_n^4}\sum^{*}\tilde{\X}_{i_1}^{'}\tilde{\X}_{i_2}\tilde{\X}_{i_3}^{'}\tilde{\X}_{i_4}+o_p(\tr(\R^2))
\end{align*}
where $\tilde{\X}_{i}=\D^{-1/2}\X_{i}$. Then, according to Theorem 2 in Chen, Zhang and Zhong (2010), we can easily obtain the result.

\subsection{Power Under Fixed Alternative}
 In this part, similar to Zhong and Chen (2011), we consider two scenarios of fixed alternatives under
\begin{align*}
\delta_{\bb}^T\bms\delta_{\bb} ~ \text{is not}~ o(1)
\end{align*}
One is
\begin{align}\label{f1}
\delta_{\bb}^T\bms \D^{-1}\bms\D^{-1}\bms\delta_{\bb}=o\left(\frac{1}{n}\delta_{\bb}^{'}\bms\delta_{\bb}\tr(\R^2)\right)
\end{align}
If $\delta_{\bb}^T\bms\delta_{\bb}$ is truly bounded, (\ref{f1}) implies $\delta_{\bb}^T\bms \D^{-1}\bms\D^{-1}\bms\delta_{\bb}=o\left(\frac{1}{n}\tr(\R^2)\right)$ which mimics the second part of (\ref{alter}).
The other is
\begin{align}\label{f2}
\frac{1}{n}\delta_{\bb}^{'}\bms\delta_{\bb}\tr(\R^2)=o\left(\delta_{\bb}^T\bms \D^{-1}\bms\D^{-1}\bms\delta_{\bb}\right)
\end{align}
If $\delta_{\bb}^T\bms\delta_{\bb}$ is truly bounded, (\ref{f1}) implies $\frac{1}{n}\tr(\R^2)=o\left(\delta_{\bb}^T\bms \D^{-1}\bms\D^{-1}\bms\delta_{\bb}\right)$ which means there is a larger discrepancies between $\bb$ and $\bb_0$.

\begin{thm}\label{th2}
Assume the condition (C1)--(C3) hold, then
\begin{itemize}
\item[(i)] under the first fixed alternatives (\ref{f1}),
\begin{align*}
\frac{n}{\sigma_{A_1}}(T_n-||\D^{-1/2}\bms (\bb-\bb_0)||^2)\cd N(0,1)
\end{align*}
where
\begin{align*}
\sigma_{A_1}^2=&2\sigma^4\tr(\R^2)+2B_1^2\tr(\R^2)+4\sigma^4\tr(\R^2)B_1\\
&+4\Delta(B_1+\sigma^2)\tr(A_1\circ A_3)+2\Delta^2\tr[(A_0\diag(A_1))^2]
\end{align*}
\item[(ii)] under the first fixed alternatives (\ref{f2}),
\begin{align*}
\frac{n}{\sigma_{A_2}}(T_n-||\D^{-1/2}\bms (\bb-\bb_0)||^2)\cd N(0,1)
\end{align*}
where
\begin{align*}
\sigma_{A_2}^2=(B_1+\sigma^2)B_3+B_2^2+\Delta\tr(A_1\circ A_2).
\end{align*}
\end{itemize}
\end{thm}
The proof of Theorem \ref{th2} is contained in a longer version of this
article. The above theorem implies that the asymptotic power of the test under the first fixed alternative (\ref{f1}) is
\begin{align*}
\beta_{T_n}(||\bb-\bb_0||)\approx\Phi\left(-\frac{\sqrt{2\tr(\R^2)}\sigma^2z_{\alpha}}{\sigma_{A_1}}+\frac{n||\D^{-1/2}\bms\delta_{\bb}||^2}{\sigma_{A_1}}\right)
\end{align*}
Note that $\sigma_{A_1}^{-1}\sqrt{2\tr(\R^2)}\sigma^2$ is always bounded from infinity because $B_1$ is not $o(1)$ and $\sigma_{A_1}^2>2B_1^2\tr(\R^2)$. When $B_1 \to \infty$, the first term converges to $0$ and then our test attains at least $50\%$ power in this case. Furthermore, if $n\sigma_{A_1}^{-1}||\D^{-1/2}\bms\delta_{\bb}||^2\to\infty$, the power will converge to $1$.
And the asymptotic power of the test under the first fixed alternative (\ref{f2}) is
\begin{align*}
\beta_{T_n}(||\bb-\bb_0||)\approx\Phi\left(-\frac{\sqrt{2\tr(\R^2)}\sigma^2z_{\alpha}}{\sqrt{n-1}\sigma_{A_2}}+\frac{n||\D^{-1/2}\bms\delta_{\bb}||^2}{\sigma_{A_2}}\right)
\end{align*}
Under fixed alternative (\ref{f2}), $\frac{1}{n}\tr{\R^2}=o(\sigma_{A_2}^2)$, which implies the first term converge to $0$. And then our test attains at least $50\%$ power in this case. Similarly, if $n\sigma_{A_2}^{-1}||\D^{-1/2}\bms\delta_{\bb}||^2\to \infty$, our test is consistent.

\section*{\small References}
{\footnotesize \baselineskip 10pt
\begin{description}
\item Anderson, T. W. (2003). {\it An Introduction to Multivariate
Statistical Analysis.} Hoboken, NJ: Wiley.

\item  Bai, Z. and Saranadasa, H. (1996). Effect of high dimension: by an
example of a two sample problem. {\it Statistica Sinica}, {\bf 6},
311--29.

\item Chen, S. X. and Qin, Y-L. (2010). A two-sample test for high-dimensional data with applications to
gene-set testing.  {\it The Annals of Statistics}, {\bf 38}, 808--835.

\item Chen, S. X., Zhang, L. -X. and Zhong, P. -S. (2010). Tests for high-dimensional covariance
matrices. {\it Journal of the American Statistical Association}, {\bf 105}, 810--815.

\item Efron, B., and Tibshirani, R. (2007). On testing the significance of sets of
genes, {\it The Annals of Applied Statistics}, {\bf 1}, 107--129.

\item Fan, J., and Lv, J. (2008). Sure independence screening for ultrahigh dimensional feature space, {\it Journal of the Royal Statistical
Society, Ser. B}, {\bf 70}, 849--911.

\item Fan, J., Hall, P., and Yao, Q. (2007). To how many simultaneous hypothesis
tests can normal student¡¯s $t$ or bootstrap calibrations be applied,
{\it Journal of the American Statistical Association}, {\bf 102}, 1282--1288.

\item Feng, L., Zou, C., Wang, Z. and Chen, B. (2013), Rank-based Score Tests for
High-Dimensional Regression Coefficients, {\it Electronic Journal of Statistics}, 7, 2131--2149.

\item Feng, L., Zou, C., Wang, Z. and Zhu, L. (2014).  Two-sample Behrens-Fisher problem for high-dimensional data, {\it Statistica Sinica}, To appear.

\item Goeman, J., Finos, L., and van Houwelingen, J. C. (2011). Testing against
a high dimensional alternative in the generalized linear model: asymptotic type I error control, {\it Biometrika}, {\bf 98}, 381--390.

\item Goeman, J., Van de Geer, S. A. and Van Houwelingen, J. C. (2006). Testing against a high-dimensional alternative.
{\it Journal of the Royal Statistical
Society, Ser. B}, {\bf 68}, 477--493.

\item Hall, P., and Heyde, C. C. (1980), {\it Martingale Limit Theory and Its Application},
New York: Academic Press.

\item Kosorok, M. R., and Ma, S. (2007). Marginal asymptotics for the ``Large p,
Small n'' paradigm: with applications to microarray data, {\it The Annals of
Statistics}, {\bf 35}, 1456--1486.

\item Lee, A. J. (1990), {\it U-Statistics: Theory and Practice}, Marcel Dekker.

\item Meinshausen, N. (2008). Hierarchical testing of variable importance, {\it Biometrika}, {\bf 95}, 265--278.

\item Newton, M., Quintana, F., Den Boon, J., Sengupta, S., and Ahlquist, P. (2007),
Random-Set Methods Identify Distinct Aspects of the Enrichment Signal
in Gene-Set Analysis, {\it The Annals of Applied Statistics}, {\bf 1}, 85--106.

\item Portnoy, S. (1984). Asymptotic behavior of the M-Estimators of $p$-regression
parameters when $p^2/n$ is large: consistency, {\it The Annals of Statistics}, {\bf 12},
1298--1309.

\item Portnoy, S. (1985). Asymptotic behavior of the M-Estimators of $p$-regression
parameters when $p^2/n$ is large: normal approximation, {\it The Annals of
Statistics}, {\bf 13}, 1403--1417.

\item Schott, J. R. (2005). Testing for complete independence in high
dimensions. {\it Biometrika} {\bf 92}, 951--956.

\item Subramanian, A., Tamayo, P., Mootha, V. K., Mukherjee, S., Ebert, B. L.,
Gillette, M. A., Paulovich, A., Pomeroy, S. L., Golub, T. R., Lander, E. S.,
and Mesirov, J. P. (2005), Gene Set Enrichment Analysis: A Knowledge-
Based Approach for Interpreting Genome-Wide Expression Profiles, {\it Proceedings
of the National Academy of Sciences}, {\bf 102}, 15545--15550.

\item Wang, H. (2009). Forward regression for ultra-high
dimensional variable screening. {\it  Journal of the American Statistical Association}. {\bf
104}, 1512--1524.

\item Zhong, P. S. and Chen, S. X. (2011). Tests for high dimensional regression coefficients with factorial designs. {\it  Journal of the American Statistical Association}, {\bf 106}, 260--274.

\end{description}}
\end{document}